# Solving the issues of multicomponent diffusion in an equiatomic NiCoFeCr medium entropy alloy

Anuj Dash, Neelamegan Esakkiraja, and Aloke Paul*
Department of Materials Engineering, Indian Institute of Science, Bengaluru 560012, India
*Corresponding author: aloke@iisc.ac.in, aloke.paul@gmail.com

**Abstract:** Estimating the diffusion coefficients experimentally in a four-component inhomogeneous alloy following the conventional diffusion couple method by intersecting three couples at the same composition is difficult unless a small composition range of constant diffusivity is identified. Additionally, the intrinsic diffusion coefficients of the components cannot be estimated in a system with more than two components. To solve these issues, we have followed the pseudo-binary and pseudo-ternary diffusion couple methods for estimating the diffusion coefficients at the equiatomic composition of NiCoFeCr medium entropy alloy. Along with the pseudo-binary interdiffusion coefficients, we have estimated the intrinsic diffusion coefficients of all the components by designing the pseudo-binary couples such that Ni and Co develop the diffusion profiles keeping Fe and Cr constant in one couple and Fe and Cr develop the diffusion profiles keeping Ni and Co constant in another couple. Subsequently, we have proposed the relations for calculating the tracer diffusion coefficients utilizing the thermodynamic details. We have found a good match with the data estimated directly following the radiotracer method at the equiatomic composition. Following, we have produced three pseudo-ternary diffusion couples intersecting at the compositions close to the equiatomic composition. The main pseudo-ternary interdiffusion coefficients of Fe are found to be higher than Ni and Co. Therefore, we have estimated different types of diffusion coefficients highlighting the complex diffusion process in the four-component NiCoFeCr medium entropy alloy.

*Keywords:* Multicomponent diffusion; Interdiffusion; High Entropy Alloys; NiCoFeCr

## 1. Introduction

The troubles with the estimation of the diffusion coefficients in a multicomponent inhomogeneous system following the diffusion couple method and lack of physical significance of the estimated data are explained in detail by DeHoff and Kulkarni [1]. On the other hand, most of the materials in applications are multicomponent in nature. Therefore, today, the diffusion studies in the multicomponent inhomogeneous material system are considered as one of the most challenging tasks.

The interdiffusion coefficients in a wide composition range of interest (not the average over a composition range) could only be estimated previously in binary and ternary inhomogeneous systems following relatively straight forward methods [1, 2]. The equations for the estimation of the interdiffusion coefficients in multicomponent systems are established based on the Onsager formalism [3, 4]. However, the number of independent equations required for the estimation of these data could not be achieved experimentally in a system with more than three components [5, 6]. This comes from the fact that estimating the interdiffusion coefficients by intersecting diffusion paths of different diffusion couples (*i.e.* by developing diffusion profiles of all the components) exactly at one composition in a single-phase region with more than three components is almost impossible. This can be handled up to a certain extent by designing the body diagonal diffusion couples in a small composition range of constant diffusivities [7] although the diffusion profiles may not intersect exactly at one particular composition (like in the ternary system). However, one may still estimate the interdiffusion coefficients at a composition close to each other [8]. This will introduce an unknown range of error depending on the system and deviation of the profiles from the composition considered for calculation. Moreover, the composition range of constant diffusivities will vary from one system to another, which need to be identified based on a set of trial experiments consisting of multiple diffusion couples. The concept behind designing a body diagonal with sufficient composition range is such that one can produce these diffusion profiles at and around the body centre but not at other parts (near edges or corners) of the multicomponent phase diagram or phase boundaries. Additionally, the intrinsic diffusion coefficients of components, which are very important for understanding the atomic mechanism of diffusion, cannot be estimated anyway in a system with more than two components. The Kirkendall marker plane should be located at the composition of intersection in all the diffusion couples, which is impossible to achieve unless found incidentally. Nevertheless, every method has its own benefits and limitations. Therefore, the possibilities of using different types of methods in different situations are equally important for understanding various aspects of multicomponent diffusion.

We circumvent the issues described above by utilizing the concept of the pseudo-binary (PB) and pseudo-ternary (PT) methods in an equiatomic NiCoFeCr medium entropy alloy. In PB diffusion couple, we can estimate the intrinsic and tracer diffusion coefficients along with the interdiffusion coefficients. Similar to the interdiffusion studies in a ternary system, we can estimate the main and cross interdiffusion coefficients by intersecting the diffusion couples exactly at one composition by producing PT diffusion profiles.

This material is common to two extensively studied High Entropy Alloys (HEA) *i.e.* NiCoFeCrMn and NiCoFeCrAl. At present, most of the diffusion studies in concentrated *i.e.* high entropy alloys is focused mainly on the possibilities of the sluggishness of diffusion rates of components. An intense discussion on this topic started after the first report by Tsai et al. [9]; however, based on the faulty analyses [10, 11]. Subsequent analysis by the tracer method [12, 13, 14] and optimization-based method in combination with Midemma's thermodynamic description [15, 16] refute this phenomenon, which was wrongly believed for some time in the absence of direct experimental evidence. At the same time, one cannot deny the importance of these material systems for developing new compositions with superior properties. Therefore, instead of just concentrating on the effect of configurational entropy on the diffusion rates of the components, it is important to develop an efficient method for studying diffusion in an inhomogeneous alloy. The interdiffusion coefficients at the composition of interest are important for a real understanding of the complicated diffusional interactions between the components.

In this article, we have estimated the interdiffusion, intrinsic and tracer diffusion coefficients at the equiatomic composition of NiCoCrFe system following the PB diffusion couple method to compare with the available tracer diffusion coefficients (measured by the radiotracer method). Therefore, this is the first study showing a comparison of data indicating the efficiency of these newly established methods in multicomponent diffusion. This also highlights the interdiffusion process in these types of couples when all the components do not develop the diffusion profiles. Following, we have produced three PT diffusion couples (keeping Cr as the constant), which intersect at compositions close to the equiatomic composition. Although these diffusion profiles can be produced in a wider composition range, we have selected a narrow composition range such that diffusion couples intersect at a composition close to the equiatomic composition, which is not easy to achieve otherwise because of serpentine nature of the diffusion profiles. As explained in section 3, these approaches help to reduce the complications for the estimation of the diffusion coefficients in a multicomponent system. Most importantly, we could intersect the diffusion couples at a particular composition (similar to the ternary system) by keeping one component as constant. Therefore, we have a unique opportunity of estimating the main and cross interdiffusion coefficients at compositions very close to the equiatomic composition of a four-component system by intersecting the diffusion couples, which is difficult otherwise. This study gives a new direction in a multicomponent medium entropy alloy for a better understanding of the complex phenomenological diffusion process.

## 2. Experimental method

The alloys for the end-member compositions of the diffusion couples were prepared by melting the pure (99.9-99.95 wt.%) components in an arc melting furnace. To ensure the compositional homogeneity, the buttons were re-melted four to five times. These were then subjected to a homogenizing heat treatment at 1200±5ºC (1473±5 K) for 50 h in high vacuum (~$10^{-4}$ Pa). Compositional analysis was carried out through spot scans at randomly selected spots in an Electron Probe Micro Analyzer (EPMA) with pure components as standards. The alloy compositions are listed in Table 1a. The end member alloys of different pseudo-binary and pseudo-ternary diffusion couples are listed in Table 1b. The differences between the desired and actual alloy compositions were found to be within 1 at.%. The deviations from the average actual compositions at different spots were found to be within ±0.1 at%. Slices with ~1 mm thickness were cut from the buttons in an Electro Discharge Machine (EDM) and prepared metallographically for flat and smooth surfaces. The flatness of the foils was maintained by holding them in a fixture, especially during the initial rough grinding process. This avoids producing the bevelled edges in the foils. The diffusion couples prepared with two dissimilar compositions were assembled in a special fixture for holding them together. These were then diffusion annealed in vacuum at 1200 °C for 50 h for the development of the interdiffusion zone. The alloys of end members of pseudo-binary and pseudo-ternary diffusion couples are listed in Table 1b. After the diffusion annealing, the samples were cross-sectioned (transverse to the interface) and prepared metallographically. Two or three diffusion profiles were then measured across the interface (using the WDS line scan mode in the EPMA). We found that all the composition profiles in a particular diffusion couple were very similar to each other. A polynomial function was used (sometimes in parts) to fit the measured diffusion profiles in Origin software. After fitting, different parts of the diffusion profiles were checked carefully to ensure a smooth fit throughout without any unreasonable deviation from the measured profile. The average error in the estimation of the diffusion coefficients is established based on the difference in diffusion coefficients from different diffusion profiles. Further details and the calculation method of the diffusion coefficients following the diffusion couple technique are described in Chapter 3 in Ref. [5].

## 3. Results and discussion

In this study, we follow the PB and PT diffusion couple technique, which extends the benefit of estimation of the data in binary and ternary systems to a multicomponent system. In a binary diffusion couple method, we can estimate the intrinsic diffusion coefficients (diffusion rates of individual components under the influence of thermodynamic driving forces) and the interdiffusion coefficients (a kind of average of the intrinsic diffusion coefficients). We can even calculate the tracer diffusion coefficients and estimate the impurity diffusion coefficients. We can draw similar benefits in a multicomponent system if we achieve a situation in which only two components develop the diffusion profiles keeping all other components as the constant throughout the interdiffusion zone. This type of diffusion couple is named as pseudo-binary (PB) diffusion couple, which was proposed for the first time by Paul, one of the authors of this article [17]. Following, Tsai et al. utilized this concept in NiCoFeCrMn high entropy alloy by naming it as the quasi-binary approach by naming it as the quasi-binary approach [9] but without following the correct steps of calculation [10]. They have shown only one ideal profile but also estimated the data even considering many non-ideal profiles without showing in the article leading to the estimation of meaningless data [11]. The estimated cross interdiffusion coefficients in a ternary system reflect on the complex diffusional interactions among the components. However, these cannot be estimated in PB diffusion couples. Therefore, as proposed by Esakkiraja and Paul one can follow a pseudo-ternary (PT) method in which only three components develop the diffusion profiles keeping all other components as the constant in the interdiffusion zone [18]. We can intersect them (like in the ternary system) if we can indeed produce ideal or near-ideal PT diffusion couples in the absence of measurable non-ideality (*i.e.* the presence of hump or uphill diffusion of the components which are supposed to remain constant). This is otherwise difficult (or impossible) in a multicomponent space even in body diagonal diffusion couples designed in a small composition range of constant diffusivities. Therefore, we need again two diffusion couples only for the estimation of the main and cross interdiffusion coefficients at the composition of an intersection after necessary modifications of the equations (as described next). These are demonstrated in a schematic diagram in Figure 1. An example of the intersection of two PT couples are shown in which three components 1, 2 and 3 develop the diffusion profiles and components 4, 5…..$n$ remain the constant. On the same diagram, examples of three PB couples are shown. For example, in PB:1-2 diffusion couple, components 1 and 2 develop the diffusion profiles keeping components 3 - $n$ as the constant. Other PB or PT couples can also be produced, such that different other components develop the diffusion profiles. However, all combinations may not develop the ideal PB or PT profiles [11]. We demonstrate the successful use of these methods to understand the diffusion phenomena in a four-component NiCoFeCr system by estimating different types of diffusion coefficients at the (equiatomic) composition of interest. Additionally, the comparison with the available tracer diffusion coefficients highlights the nature of the diffusion coefficients estimated following these methods.

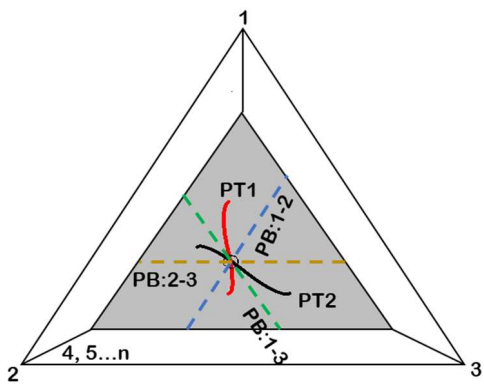

Figure 1 The examples of pseudo-binary and pseudo-ternary diffusion couples in a multicomponent system.

## 3.1 The Pseudo-Binary (PB) diffusion couples in NiCoFeCr system

We first estimate the interdiffusion coefficients following the PB method. We have possibilities of producing six PB diffusion couples making two components to develop diffusion profiles in different combinations in a four-component system. However, all the couples may not produce the ideal PB diffusion profiles such that only two components develop the diffusion profiles keeping other components constant throughout the interdiffusion zone. In a non-ideal diffusion couple, the components which are kept with the same composition in both the end-members may develop a hump or uphill diffusion. In such a situation, this method cannot be utilized unless the non-ideality is minor [10, 11, 19]. Tsai et al. utilized this method (although named it differently as the quasi-binary approach) [9] even in the presence of major non-ideality without showing the profiles except one ideal profile [20] and also without utilizing a proper step leading to the estimation of the data without any physical significance [11]. They wrongly defended their approach not based on valid logic but by citing one of our articles showing a PB profile with undulations resulted because of error in measurements without showing the presence of a major hump [20]. In this article, we aim at discussing the diffusion phenomenon by considering the ideal diffusion profiles and, therefore, we have produced only the PB Ni-Co(fixed Fe, Cr) and Fe-Cr(fixed Ni, Co) diffusion couples based on our preliminary assessment. More importantly, these two couples allow estimating the intrinsic and tracer diffusion coefficients of all the components.

Table 1 (a) The designation of alloys, the desired and actual compositions and (b) the alloys used for preparing the PB and PT diffusion couples.

| Alloy designation | Desired Compositions | Actual Compositions |
|---|---|---|
| A | $Ni_{30}Co_{20}Fe_{25}Cr_{25}$ | $Ni_{29.4}Co_{19.8}Fe_{25.1}Cr_{25.7}$ |
| B | $Ni_{20}Co_{30}Fe_{25}Cr_{25}$ | $Ni_{19.5}Co_{29.7}Fe_{25.1}Cr_{25.7}$ |
| C | $Ni_{25}Co_{25}Fe_{20}Cr_{30}$ | $Ni_{24.6}Co_{24.7}Fe_{20}Cr_{30.7}$ |
| D | $Ni_{25}Co_{25}Fe_{30}Cr_{20}$ | $Ni_{24.6}Co_{24.7}Fe_{30.1}Cr_{20.6}$ |
| E | $Ni_{27.5}Co_{27.5}Fe_{20}Cr_{25}$ | $Ni_{26.9}Co_{27.2}Fe_{20.2}Cr_{25.7}$ |
| F | $Ni_{20}Co_{20}Fe_{35}Cr_{25}$ | $Ni_{19.4}Co_{19.8}Fe_{35.1}Cr_{25.7}$ |
| G | $Ni_{20}Co_{27.5}Fe_{27.5}Cr_{25}$ | $Ni_{19.3}Co_{27.3}Fe_{27.7}Cr_{25.7}$ |
| H | $Ni_{35}Co_{20}Fe_{20}Cr_{25}$ | $Ni_{34.3}Co_{19.8}Fe_{20.2}Cr_{25.7}$ |
| I | $Ni_{27.5}Co_{20}Fe_{27.5}Cr_{25}$ | $Ni_{26.9}Co_{19.8}Fe_{27.6}Cr_{25.7}$ |
| J | $Ni_{20}Co_{35}Fe_{20}Cr_{25}$ | $Ni_{19.5}Co_{34.5}Fe_{20.3}Cr_{25.7}$ |

(a)

| Diffusion couples | End Member - 1 | End Member - 2 |
|---|---|---|
| Pseudo binary(PB) - 1 | A | B |
| Pseudo binary (PB) - 2 | C | D |
| Pseudo ternary (PT) - 1 | E | F |
| Pseudo ternary (PT) - 2 | G | H |
| Pseudo ternary (PT) - 3 | I | J |

(b)

Since we are interested in estimating the data at (or near) the equiatomic composition, two diffusion couples are produced by coupling

Pseudo-Binary Diffusion couple (PB) -1: Alloy A *i.e.* $Ni_{30}$-$Co_{20}(Fe_{25}Cr_{25})$ and Alloy B *i.e.* $Ni_{20}$-$Co_{30}(Fe_{25}Cr_{25})$.

Pseudo-Binary Diffusion couple (PB) -2: Alloy C *i.e.* $(Ni_{25}Co_{25})Fe_{30}$-$Cr_{20}$ and Alloy D *i.e.* $(Ni_{25}Co_{25})Fe_{20}$-$Cr_{30}$.

These experiments are conducted at 1200 °C for 50 h. As shown in Figure 2a and b, the diffusion couples are produced such that Ni and Co develop the diffusion profiles keeping Fe and Cr composition fixed in PB-1. In PB-2, Fe and Cr develop the diffusion profiles keeping Ni and Co composition fixed. Most importantly, both the PB couples produce the ideal diffusion profiles such that the components which are kept at a fixed value in both the end members of a diffusion couple remain constant without producing a measurable hump in the interdiffusion zone. The PB interdiffusion flux estimated concerning one of the diffusing components *i* and the PB interdiffusion coefficient considering a constant molar volume are related by [19]

$$V_m \tilde{J}_i = -\tilde{D} \frac{dM_i}{dx} \qquad (1a)$$

$$\tilde{J}(Y^*_{M_i}) = -\frac{M_i^+ - M_i^-}{2tV_m}\left[(1-Y^*_{M_i})\int_{x-\infty}^{x^*} Y_{M_i}\,dx + Y^*_{M_i}\int_{x^*}^{x+\infty}(1-Y_{M_i})\,dx\right] \quad (1b)$$

$$Y_{M_i} = \frac{M_i - M_i^-}{M_i^+ - M_i^-} \quad (1c)$$

$Y_{M_i}$ is the modified composition normalized variable, where $M_i = \frac{N_i}{N_v} = \frac{N_i}{1-N_f}$. $N_v = N_1 + N_2$ is the sum of the composition of elements which develop the diffusion profiles and $N_f = N_3 + N_4 \ldots \ldots N_n$ is the sum of the composition of elements which remain constant in an interdiffusion zone. $x$ is the position parameter, and $t$ is the annealing time. This fulfils the condition such that the sum of the interdiffusion fluxes i.e. $(\tilde{J}_1 + \tilde{J}_2) = 0$. The components which are kept constant take part in the diffusion process locally for redistribution of components; however, do not develop the diffusion profiles and therefore do not contribute to the interdiffusion flux.

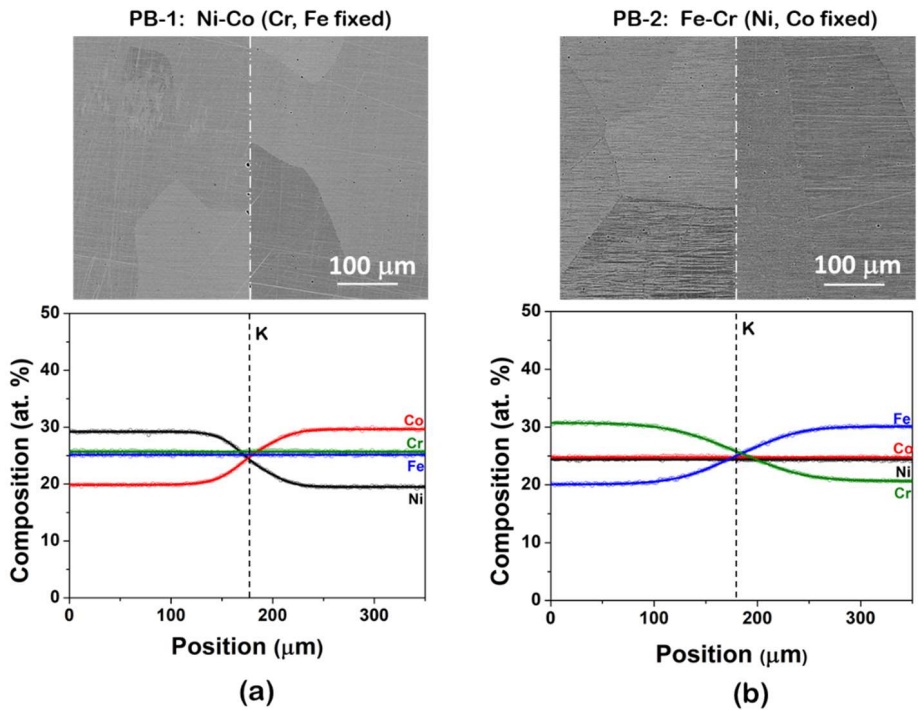

Figure 2 The microstructure showing the locations of the Kirkendall marker planes and the composition profiles of the pseudo-binary (PB) diffusion couples annealed at 1200 °C for 50 h: (a) Ni$_{30}$Co$_{20}$(Fe$_{25}$Cr$_{25}$) - Ni$_{20}$Co$_{30}$(Fe$_{25}$Cr$_{25}$) (b) Fe$_{20}$Cr$_{30}$(Ni$_{25}$Co$_{25}$) - Fe$_{30}$Cr$_{20}$(Ni$_{25}$Co$_{25}$).

By relating Equation 1a and b, the PB interdiffusion coefficients can be directly estimated from the measured composition profile using

$$\tilde{D} = \frac{1}{2t}\left(\frac{dx}{dY_{M_i}}\right)_{x^*}\left[(1-Y^*_{M_i})\int_{x-\infty}^{x^*} Y_{M_i}\,dx + Y^*_{M_i}\int_{x^*}^{x+\infty}(1-Y_{M_i})\,dx\right] \quad (2)$$

It can be realized from Equation 2 that one can estimate the interdiffusion coefficients by using the composition profiles directly i.e. without modifying the composition profile since all the parts are expressed with respect to $Y_{M_i} = \frac{M_i - M_i^-}{M_i^+ - M_i^-} = \frac{N_i - N_i^-}{N_i^+ - N_i^-} = Y_{N_i}$, where $Y_{N_i}$ is the composition normalized variable proposed by Sauer-Freise [21]. However, one should use the modified composition profiles for the calculation of PB interdiffusion flux. This is also important in the equation which relates the PB interdiffusion coefficient with the intrinsic diffusion coefficients and tracer diffusion coefficients considering the thermodynamic driving forces and the vacancy wind effect (as discussed next). For the sake of avoiding confusion, we have used the modified composition profiles throughout. This is rather logical considering the diffusion process in these types of diffusion couples in which all the components do not contribute to the development of the interdiffusion profile.

The intrinsic diffusion coefficients can be estimated at the Kirkendall marker plane utilizing [19]

$$D_i = \frac{1}{2t}\left(\frac{\partial x}{\partial M_i}\right)_K \left[M_i^+ \int_{x-\infty}^{x^K} Y_{M_2}\, dx - M_i^- \int_{x^K}^{x+\infty} (1 - Y_{M_2})\, dx\right] \quad (3)$$

Now we compare the PB interdiffusion profiles with the tracer diffusion coefficients reported by Vaidya et al. [12]. As shown in Figure 3a, we have replotted their data and extended up to the experimental temperature of this study *i.e.* 1200 °C (1/T = 6.79×10$^{-4}$ K$^{-1}$). Ni and Co have lower diffusion rates compared to Fe and Cr. Therefore, we should expect a smaller interdiffusion zone length in PB-1 in which Co and Ni develop the diffusion profiles compared to the interdiffusion zone length in the PB-2 in which Fe and Cr develop the diffusion profiles. This is indeed seen to be the case as is clear from the diffusion profiles plotted in Figures 2a and b. The same differences are also reflected in the estimated PB interdiffusion coefficients, as plotted in Figure 3b. The intrinsic diffusion coefficients are estimated at the Kirkendall marker plane identified by the microstructural evolution, as shown in Figure 2. This is demarcated by duplex morphology based on the location of the marker plane [22]. The estimated PB interdiffusion coefficients and the intrinsic diffusion coefficients at the marker plane are listed in Table 2a. These are related by [19]

$$\widetilde{D} = M_2 D_1 + M_1 D_2 \quad (4)$$

It should be noted here that the relation above is correctly expressed with the modified composition (M), which can be derived considering the interdiffusion process in PB diffusion couples. Confusion has created concerning the reference Manning factor because of the use of composition (N) instead of Modified composition [23], which might have resulted unknowingly because of a typing error in Ref. [24]. We have advocated the use of M instead of N in this type of couple from the beginning [10, 19].

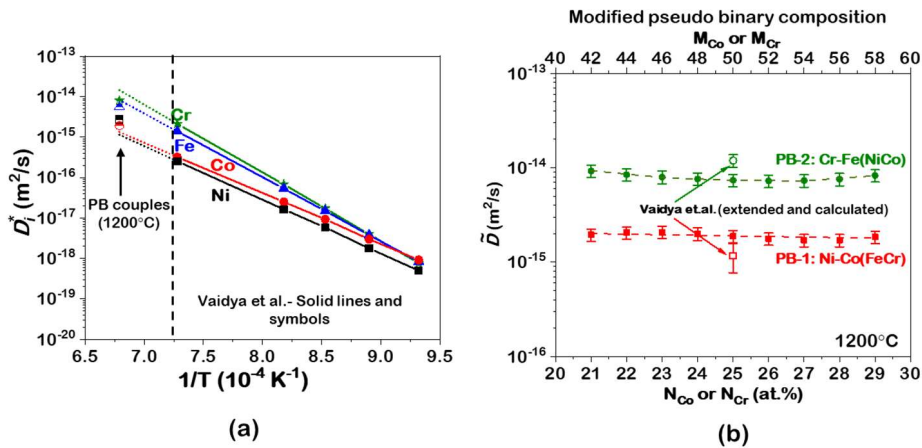

Figure 3 (a) The estimated tracer diffusion coefficients from the pseudo-binary diffusion couples are compared with the data reported by Vaidya et al. [19] following the radio-tracer method (b) The estimated interdiffusion coefficients in this study are compared with the values calculated utilizing the tracer diffusion coefficients reported by Vaidya et al. These are estimated at 1200°C.

One can even calculate the tracer diffusion coefficients ($D_i^*$) with the intrinsic diffusion coefficient for a constant molar volume by

$$D_1 = D_1^* \emptyset_{11}(1 + W_1) \tag{5a}$$

$$D_2 = D_2^* \emptyset_{22}(1 - W_2) \tag{5b}$$

Where $\emptyset_{ii} = \frac{dlna_i}{dlnN_i}$ is the main thermodynamic factor of component $i$. These are calculated from activity of components $a_i$ ThermoCalc Software using High Entropy Alloy (HEA) Database. These values are listed in Table 2a. The vacancy wind effect concerning the modified compositions in a PB couple can be expressed as

$$1 + W_1 \cong 1 + \frac{2M_1\left(D_1^* - D_2^*\frac{\emptyset_{22}}{\emptyset_{11}}\right)}{S_o(M_A D_A^* + M_B D_B^*)}; \quad 1 - W_2 \cong 1 - \frac{2M_2\left(D_1^*\frac{\emptyset_{11}}{\emptyset_{22}} - D_2^*\right)}{S_o(M_A D_A^* + M_B D_B^*)}; \tag{5c}$$

$S_o$ is a constant, which depends on the crystal structure, for example, 7.15 in the FCC phase. This equation is similar to the relation proposed in the binary system by Manning expressed with respect to composition ($N_i$) [25]. In a PB couple, this is expressed with respect to the modified composition ($M_i$) and relevant thermodynamic parameters. In a binary system, the thermodynamic parameters of both the components are the same, M converts to N and these relations transform to the equation proposed by Manning in a binary system. An approximate sign is used because of the assumptions considered for the utilization of these relations in near ideal diffusion profiles, which will be found frequently in many couples. These methods should not be used in the presence of major hump i.e. major non-ideality for the estimation of meaningless data. The calculated tracer diffusion coefficients are listed in Table 2a. As mentioned in Table 2a and 2b, these are not estimated exactly at the equiatomic composition. However, there is not much difference in the PB interdiffusion coefficients at the equiatomic composition and the compositions of the Kirkendall marker planes at which these values are calculated. Therefore, we can safely compare these values with the tracer diffusion coefficients estimated directly following radiotracer method by Vaidya et al. [12], as shown in Figure 3a. The contribution from the vacancy wind effect is found to be negligible ($W_i \leq 0.05$) in the PB couples considered in this study. Since the value of the vacancy wind effect is anyway very small in this type of system, one can even use the relation proposed by Manning in the binary system for the sake of simplicity by calculating an approximate value without making a significant error, as it is done in the NiCoCr system [26]. However, the intrinsic/interdiffusion and tracer diffusion coefficients should be always related to the thermodynamic parameter of the component of interest instead of considering the same value for both the components. The vacancy wind effect may remain a topic of interest for a more accurate relation considering different types of a non-ideal situation. This will be discussed again in future based on extensive analysis. However, indeed, this will not make a significant contribution in most of the PB and PT couples in various systems in comparison to the error introduced in the experimentally estimated data similar to the situation in conventional binary and ternary diffusion couples. Therefore, one may safely ignore it for the sake of avoiding complications of equations without making any mentionable difference.

Similarly, we can estimate the PB interdiffusion coefficients at the equiatomic composition utilizing the tracer diffusion coefficients reported by them and compare it with the interdiffusion coefficients estimated at the equiatomic composition in this study. The PB interdiffusion and tracer diffusion coefficients with respect to the modified composition profile are related by

$$\widetilde{D} = (M_2 D_1^* \emptyset_{11} + M_1 D_2^* \emptyset_{22})W_{12} \tag{6a}$$

At the equiatomic composition of these couples, we have $M_i = \frac{N_i}{N_v} = \frac{0.25}{0.5} = 0.5$ for both the diffusing components. The vacancy wind effect ($W_{12}$) of PB interdiffusion with respect to the modified composition in a PB couple can be expressed as

$$W_{12} \cong 1 + \frac{2M_1 M_2 (D_1^* - D_2^*)(D_1^* \emptyset_{11} - D_2^* \emptyset_{22})}{S_o (M_1 D_2^* \emptyset_{22} + M_2 D_1^* \emptyset_{11})(M_1 D_1^* + M_2 D_2^*)} \tag{6b}$$

An approximate sign is used in the relation of the vacancy wind effect for the use of this relation in near ideal PB diffusion profiles. Again, in a binary system, the thermodynamic parameters of both components are the same. M is converted to N and this relation transforms to the relation proposed by Manning in a binary system. It also should be noted here that (unlike the binary system), the thermodynamic parameters should not be calculated from the interdiffusion coefficients using the tracer diffusion coefficients in a PB couple [23, 24], since these are not the same for both the components. Rather, these should be estimated considering the intrinsic diffusion coefficients, as expressed in Equation 5.

Table 2 (a) Estimated interdiffusion and intrinsic diffusion coefficients at the Kirkendall marker plane at 1200°C. Following, the tracer diffusion coefficients are calculated utilizing the thermodynamic details. (b) Reported tracer diffusion coefficients by Vaidya et al., calculated interdiffusion coefficients using these values and the estimated interdiffusion coefficients in this study.

|  | Composition (%) (K Plane) | Interdiffusion coefficients (×10⁻¹⁵ m²/s) | Intrinsic diffusion coefficients (×10⁻¹⁵ m²/s) | | Tracer diffusion coefficients (calculated, this study) (×10⁻¹⁵ m²/s) | | Thermodynamic factor | |
|---|---|---|---|---|---|---|---|---|
| PB-1 Ni-Co (Cr, Fe fixed) | $M_{Co}$ = 50.7 i.e. ($N_{Ni}$ = 24.1, $N_{Co}$ = 24.9) (fixed $N_{Fe}$ = 25.2, $N_{Cr}$ = 25.8) | $\widetilde{D}_{Ni-Co}$ = 2.2±0.3 | $D_{Ni}$ = 2.5±0.4 | $D_{Co}$ = 1.8±0.3 | $D_{Ni}^*$ = 2.7±0.4 | $D_{Co}^*$ = 1.9±0.28 | $\emptyset_{NiNi}$ 0.88 | $\emptyset_{CoCo}$ 1.01 |
| PB-2 Fe-Cr (Co, Ni fixed) | $M_{Cr}$ = 51 i.e. ($N_{Fe}$=24.9, $N_{Cr}$ = 25.9) (fixed $N_{Ni}$ = 24.4, $N_{Co}$ = 24.8) | $\widetilde{D}_{Fe-Cr}$ = 7.3±1 | $D_{Fe}$ = 5.2±0.8 | $D_{Cr}$ = 9.5±2 | $D_{Fe}^*$= 5.6±0.8 | $D_{Cr}^*$ = 8±1.2 | $\emptyset_{FeFe}$ 0.98 | $\emptyset_{FeFe}$ 1.14 |

(a)

| Vaidya et al. (equiatomic composition) | | This study, estimated at the equiatomic composition |
|---|---|---|
| Tracer diffusion coefficients (x10⁻¹⁵ m²/s) | Interdiffusion coefficients (x10⁻¹⁵ m²/s) (calculated, this study) | Interdiffusion coefficients (x10⁻¹⁵ m²/s) |
| $D_{Ni}^*$ = 1.2±0.11 | $\widetilde{D}_{Ni-Co}$ = 1.2±0.2 | $\widetilde{D}_{Ni-Co}$ = 1.9 ±0.29 |
| $D_{Co}^*$ = 1.4±0.05 | | |
| $D_{Fe}^*$ = 8.7±0.08 | $\widetilde{D}_{Fe-Cr}$ = 12.5±1.9 | $\widetilde{D}_{Fe-Cr}$ = 7.9 ±1.2 |
| $D_{Cr}^*$ = 14. 3±0.28 | | |

(b)

As shown in Figure 3a, the tracer diffusion coefficients reported by Vaidya et al. [12] are extended to 1200 °C. The calculated PB interdiffusion coefficients following Equation 6a are included in Figure 3b for the comparison of the data estimated in this study. There is a good match of the data with a small difference. Dąbrowa et al. [15] produced the diffusion couples experimentally but estimated the data following an optimization-based method combined with Midemma's thermodynamic description. We found a reasonably good match with the data of Fe and Cr tracer diffusion coefficients. They have stated that an error is introduced in their calculation of the slowest diffusing species, *i.e.* Ni, because of the inherent problem associated with the optimization-based method. However, we found a high difference in the data of Co.

It should be noted at this point that the actual error in thermodynamics parameters extracted from ThermoCalc is not known. Moreover, the data estimated by completely different methods (diffusion couple technique and radiotracer method) are compared. Each method has its own possibilities for errors. Even the sample preparation and the use of different furnaces (might be with different rates of heating) may introduce different types of errors, which will not necessarily be reflected in the error range but on the average values of the data. For example, even higher differences are reported in binary β-NiAl phase [27]. A consideration of constant molar volume (since variation of the lattice parameters with composition in a multicomponent system is not known) also includes a certain level of error. Considering all these factors and comparing the differences in data reported between the two different groups, we can confidently state that there is a very good match between the data generated following two different methods by different groups.

### 3.2 The Pseudo-Ternary (PT) diffusion couples in NiCoFeCr system

As explained in the beginning, instead of following the conventional method by producing diffusion profiles of all the components, (which was followed during the last several decades) we follow the PT method by keeping a component (Cr) the constant in the quaternary NiCoFeCr system. In such a situation, we have $\tilde{J}_4 = \tilde{J}_{Cr} = 0$ and $\frac{dN_4}{dx} = \frac{dN_{Cr}}{dx} = 0$ in the interdiffusion zone. Therefore, with respect to the modified compositions, the interdiffusion fluxes and interdiffusion coefficients are related by [19]

$$V_m\tilde{J}_1 = -\tilde{D}_{11}^3 \frac{dM_1}{dx} - \tilde{D}_{12}^3 \frac{dM_2}{dx} \tag{7a}$$

$$V_m\tilde{J}_2 = -\tilde{D}_{21}^3 \frac{dM_1}{dx} - \tilde{D}_{22}^3 \frac{dM_2}{dx} \tag{7b}$$

$$\tilde{J}_3 = -(\tilde{J}_1 + \tilde{J}_2) \tag{7c}$$

$\tilde{D}_{ii}$, the main PT interdiffusion coefficients of component *i*, are related to the composition gradient of the same component. $\tilde{D}_{ij}$, the cross PT interdiffusion coefficients of component *i*, are related to the composition gradient of another component *j*. These are written with respect to the modified composition ($M_i$) profiles since only three components are involved in producing the diffusion profiles although four components are present in the alloy. This can be expressed as $M_i = \frac{N_i}{N_v} = \frac{N_i}{1-N_f}$, $N_v = N_1 + N_2 + N_3$ and $N_f = N_4$ [19]. Note that the modified compositions in a pseudo-ternary system are different than the modified compositions in a pseudo-binary system. Therefore, there are four PT interdiffusion coefficients to be determined. Since two equations can be written from one diffusion couple, we need two diffusion couples to intersect at one composition. This decreases the complication to a great extent. If one component indeed remains as the constant in a four-component system, we do not have difficulties to intersect two diffusion couples (a situation similar to the ternary system, see Figure 1) since the composition profiles are restricted in two-dimension instead of a three-dimensional space. The PT interdiffusion fluxes can be estimated utilizing Equation 1b.

Based on our pre-assessment, we expected to produce at least two ideal PT profiles out of the three possibilities examined in this study in the NiCoFeCr system by keeping Cr with a fixed composition. These are indeed found, but in all the three diffusion couples, as shown in Figure 4. These experiments are conducted at 1200 °C for 50 h. To realize the diffusion paths, the modified composition profiles are plotted on Gibb's triangle, as shown in Figure 5. Since Cr is kept as the constant, we targeted to melt the alloys with 25 at.% in all the end members. However, actually, we have the composition of 25.7 at.% Cr in the melted alloys, which is considered for the estimation of the diffusion coefficients. Most importantly, we have the same average composition of this element in all the terminal alloys. This is not a difficult task to achieve after a few repeated efforts of remelting. This is a basic requirement of producing good quality PT couples. We decided to produce three PT couples so that there is a higher chance of finding two ideal or nearly ideal PT couples, which is the minimum requirement of estimating the diffusion coefficients from PT couples. Moreover, the end member compositions of these couples are chosen such that there is a similar angle between the lines connecting the composition of end members of a diffusion couple on Gibb's triangle. The serpentine nature is very prominent in PT–2 and PT–3, whereas, this is less prominent in PT–1 couple. Because of this serpentine nature, these couples intersect at a different composition (circled by a solid line) compared to the intersection of lines connecting the end members of the diffusion couples (circled by a dotted line). However, all three couples have intersected at very close compositions (within 1 at.% of the equiatomic composition), which is shown in the enlarged plot. The modified and actual compositions of the intersections are also mentioned. One can improve it further by changing one of the end-member composition of a particular couple. Better control of the diffusion couple path is expected in PT-1 because of less prominent serpentine nature. For example, this couple is prepared by coupling alloys E and F. One may get even closer intersection composition of the diffusion couples with a small change in composition E (towards higher Co). However, we did not do this exercise because of very similar values estimated from different combinations in this small composition range of the intersections, which is explained next.

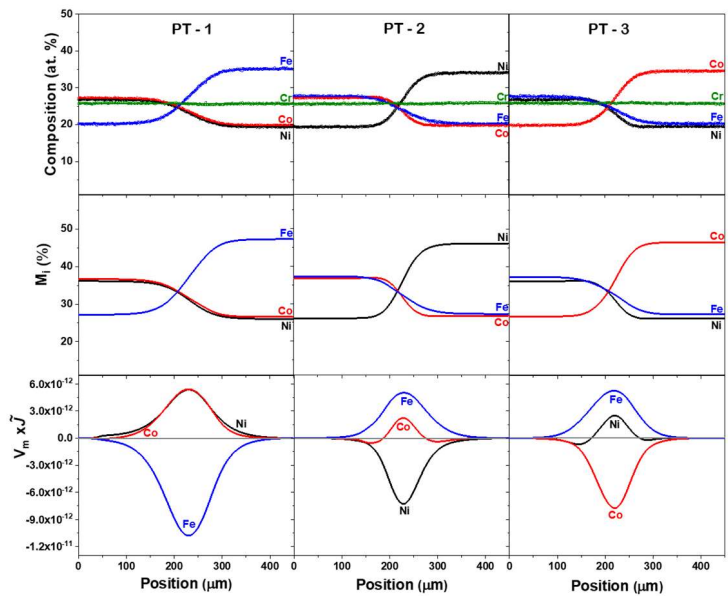

figure 4 Composition, modified composition profiles and volume flux of the three pseudo-ternary (PT) diffusion couples: PT-1 $Ni_{27.5}Co_{27.5}Fe_{20}(Cr_{25})$ - $Ni_{20}Co_{20}Fe_{35}(Cr_{25})$, PT-2 $Ni_{20}Co_{27.5}Fe_{27.5}(Cr_{30})$ - $Ni_{35}Co_{20}Fe_{20}(Cr_{25})$ and PT-3 $Ni_{27.5}Co_{20}Fe_{27.5}(Cr_{30})$ - $Ni_{20}Co_{35}Fe_{20}(Cr_{25})$. These are developed at 1200°C after annealing for 50 h.

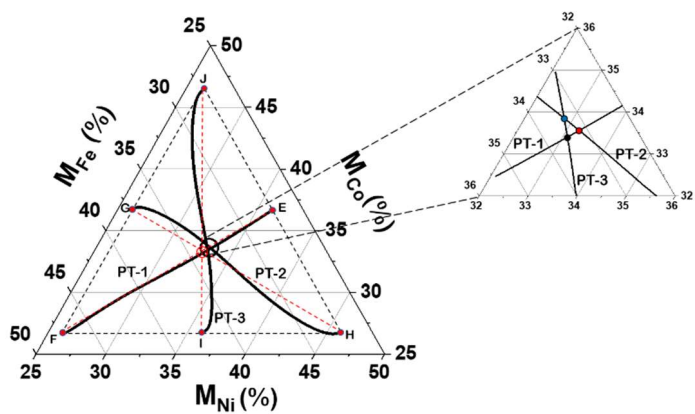

Figure 5 Modified composition profiles of pseudo-ternary diffusion couples (PT) on Gibb's triangle: PT-1 $Ni_{27.5}Co_{27.5}Fe_{20}(Cr_{25})$ - $Ni_{20}Co_{20}Fe_{35}(Cr_{25})$, PT-2 $Ni_{20}Co_{27.5}Fe_{27.5}(Cr_{25})$ - $Ni_{35}Co_{20}Fe_{20}(Cr_{25})$ and PT-3 $Ni_{27.5}Co_{20}Fe_{27.5}(Cr_{25})$ - $Ni_{20}Co_{35}Fe_{20}(Cr_{25})$. The cross of different couples can be realized in the enlarged part. These are developed at 1200°C after annealing for 50 hrs.

Since we have three profiles intersecting (almost) at compositions close to the equiatomic composition, we can estimate three different sets of data by considering different combinations of the diffusion couples (PT-1 and PT-2, PT-1 and PT-3 and PT-2 and PT-3), which are listed in Table 3. It should be noted here that we have a very good match of the data considering different combinations. Moreover, we can estimate the data considering different component as the dependent variable [2]. For example, the PT interdiffusion coefficients with Fe as the dependent variable are estimated utilizing the interdiffusion fluxes calculated from the composition profiles of Ni and Co. Similarly, the PT interdiffusion coefficients with Ni as the dependent variable considering the interdiffusion fluxes of Co and Fe and the interdiffusion coefficients with Co as the dependent variable considering the interdiffusion fluxes of Ni and Fe are estimated. As expected, the interdiffusion coefficients estimated considering different component as the dependent variable are found to be related between themselves following the equations given in Ref. [2] for a ternary system. The main PT interdiffusion coefficients of Fe have higher values compared to the main PT interdiffusion coefficients of Ni and Co in this alloy, which falls in the line of estimated intrinsic and tracer diffusion coefficients estimated following the PB method (this study) and reported tracer diffusion coefficients estimated by the radiotracer method (Vaidya et al. [12]).

Table 3 (a) The composition at the cross of different diffusion couples and (b) estimated main and cross interdiffusion coefficients considering different components as the dependent variable at 1200°C.

**Cross-section compositions of the PT couples**

| Cross of PT couples | Modified composition at the cross ($M_i$ %) | Composition at the cross ($N_i$ %) |
|---|---|---|
| PT-1 / PT-2 | $M_{Ni}$ = 33.2, $M_{Co}$ = 33.6, $M_{Fe}$ = 33.2 | $N_{Ni}$ = 24.7, $N_{Co}$ = 24.9, $N_{Fe}$ = 24.7, (fixed $N_{Cr}$ = 25.7) |
| PT-1 / PT-3 | $M_{Ni}$ = 33.1, $M_{Co}$ = 33.4, $M_{Fe}$ = 33.5 | $N_{Ni}$ = 24.6, $N_{Co}$ = 24.8, $N_{Fe}$ = 24.9, (fixed $N_{Cr}$ = 25.7) |
| PT-2 / PT-3 | $M_{Ni}$ = 32.8, $M_{Co}$ = 33.9, $M_{Fe}$ = 33.3 | $N_{Ni}$ = 24.4, $N_{Co}$ = 25.2, $N_{Fe}$ = 24.7, (fixed $N_{Cr}$ = 25.7) |

(a)

**Fe dependent variable**

| $\widetilde{D}^{Fe}_{CoCo}$ (×10⁻¹⁵ m²/s) | $\widetilde{D}^{Fe}_{CoNi}$ (×10⁻¹⁵ m²/s) | $\widetilde{D}^{Fe}_{NiNi}$ (×10⁻¹⁵ m²/s) | $\widetilde{D}^{Fe}_{NiCo}$ (×10⁻¹⁵ m²/s) | |
|---|---|---|---|---|
| 4.6±0.7 | 2.2±0.3 | 4.3±0.6 | 2.3±0.3 | PT-1 / PT-2 |
| 4.7±0.7 | 1.7±0.2 | 4.5±0.7 | 1.8±0.3 | PT-1 / PT-3 |
| 5.3±0.8 | 2.6±0.4 | 3.8±0.5 | 1.4±0.2 | PT-2 / PT-3 |

**Ni dependent variable**

| $\widetilde{D}^{Ni}_{CoCo}$ (×10⁻¹⁵ m²/s) | $\widetilde{D}^{Ni}_{CoFe}$ (×10⁻¹⁵ m²/s) | $\widetilde{D}^{Ni}_{FeFe}$ (×10⁻¹⁵ m²/s) | $\widetilde{D}^{Ni}_{FeCo}$ (×10⁻¹⁵ m²/s) | |
|---|---|---|---|---|
| 2.4±0.3 | -2.2±0.3 | 6.5±1 | -0.4±0.06 | PT-1 / PT-2 |
| 3.0±0.4 | -1.7±0.2 | 6.2±0.9 | -0.4±0.05 | PT-1 / PT-3 |
| 2.7±0.4 | -2.6±0.4 | 6.4±0.9 | -0.3±0.04 | PT-2 / PT-3 |

**Co dependent variable**

| $\widetilde{D}^{Co}_{FeFe}$ (×10⁻¹⁵ m²/s) | $\widetilde{D}^{Co}_{FeNi}$ (×10⁻¹⁵ m²/s) | $\widetilde{D}^{Co}_{NiNi}$ (×10⁻¹⁵ m²/s) | $\widetilde{D}^{Co}_{NiFe}$ (×10⁻¹⁵ m²/s) | |
|---|---|---|---|---|
| 6.9±0.92 | 0.4±0.06 | 2.0±0.3 | -2.3±0.3 | PT-1 / PT-2 |
| 6.5±0.91 | 0.4±0.05 | 2.7±0.4 | -1.8±0.3 | PT-1 / PT-3 |
| 6.7±0.92 | 0.3±0.04 | 2.4±0.3 | -1.4±0.2 | PT-2 / PT-3 |

(b)

Following Equation 7, the increase or decrease of the interdiffusion flux depends both on the sign of the cross interdiffusion coefficient and the concentration gradient based on the diffusion direction of the components. The positive diffusional interaction of component $j$ with component $i$ leads to enhancement of the interdiffusion flux of component $i$ down the concentration gradient and reduction up the concentration gradient of component $j$. The opposite is true for the negative diffusional interaction [8]. These can be witnessed in different PT diffusion couples in comparison to the estimated cross interdiffusion coefficients and direction of diffusing components. For example, the interdiffusion fluxes of both Ni and Co are enhanced in PT-1 since these components diffuse in the same direction with positive cross interdiffusion coefficients ($\widetilde{D}^{Fe}_{CoNi}$ and $\widetilde{D}^{Fe}_{NiCo}$) as listed in Table 3b. Additionally, there is a similarity in the diffusion profiles of Co and Ni in this, since Co and Ni diffuse in the same direction but opposite to Fe. This is reflected in the PT interdiffusion fluxes of these two components with a small difference, as shown in Figure 4. This is the reason for finding a less prominent serpentine path on modified Gibb's triangle, as shown in Figure 5.

On the other hand, there is a similarity in PT-2 and PT-3 diffusion profiles. In PT-2, Co and Fe diffuse opposite to Ni and in PT-3, Ni and Fe diffuse opposite to Co. As shown in Figure 4, the PT interdiffusion flux of Co in PT-2 and the PT interdiffusion flux of Ni in PT-3 changes the sign. It means these components go through the zero (interdiffusion) flux plane. These characteristics are reflected with more prominent serpentine nature in the diffusion paths of these diffusion couples. Most importantly, we could estimate the PT interdiffusion coefficients at the composition of interest (close to the equiatomic composition) in an important NiCoCrFe alloy highlighting diffusional interactions between components.

The interdiffusion coefficients estimated following the PB and PT methods can be compared. The concept of the PB couple is established by removing the contributions from the cross-terms and therefore forcing the main interdiffusion coefficients to an equal value. In a PB diffusion couple of components 1 and 2, we have $\tilde{J}_1 = -\tilde{D}(1)\frac{dM_1}{dx}$ and $\tilde{J}_2 = -\tilde{D}(2)\frac{dM_2}{dx}$. At the same time, we have $dM_1 + dM_2 = 0$ and $\tilde{J}_1 + \tilde{J}_2 = 0$. Therefore, we have $\tilde{D}(1) = \tilde{D}(2)$. If we now consider the PB Ni-Co diffusion couple, in which Fe and Cr are kept as the constant, the contributions from the cross-terms are removed. Therefore, the interdiffusion flux will change such a way that $\tilde{D}_{NiNi}^{Co}$ and $\tilde{D}_{CoCo}^{Ni}$ will change to the same value of PB $\tilde{D}(Ni) = \tilde{D}(Co)$. It can be seen that the PB Ni-Co interdiffusion coefficient at the equiatomic composition has a value close to the values of $\tilde{D}_{NiNi}^{Co}$ and $\tilde{D}_{CoCo}^{Ni}$. At present, we are doing further analysis to examine how exactly the interdiffusion coefficients estimated in PB and PT diffusion couples are related, which will be published in future. The analyses in this article demonstrate the strength of an efficient and relatively easy to handle experimental approach for understanding the complicated diffusion process in a multicomponent system.

## 4. Conclusion

In this study, we have chosen NiCoFeCr four-component medium entropy alloy to highlight different aspects of multicomponent diffusion utilizing the recently proposed PB and PT diffusion couple methods. For the sake of comparison with the tracer diffusion coefficients, we have first estimated these parameters of all the components indirectly from the PB diffusion couple experiments utilizing the thermodynamic parameters. Following, we extended our analysis for the estimation of the main and cross interdiffusion coefficients in PT diffusion couples highlighting diffusional interactions between the components. The outcome of this study can be summarized as

- Two PB diffusion couples are produced in the NiCoFeCr four-component system such that the diffusion rates of all the four components can be estimated. Reported tracer diffusion coefficients indicate that Fe and Cr have higher diffusion rates compared to Ni and Co. Reflecting this, the interdiffusion zone thickness of the PB Fe-Cr (fixed Ni, Co) couple is found to be higher than PB Ni-Co(fixed Fe, Cr) couple (Figure 2).
- The Kirkendall marker plane in the PB couples is identified based on the presence of duplex microstructure for the estimation of the intrinsic diffusion coefficients. Following, the relations to estimate the tracer diffusion coefficients utilizing the thermodynamic parameters are proposed. A very good match is found with the data measured by the radiotracer method.
- The main PT interdiffusion coefficients estimated at the equimolar composition from the PT diffusion couples indicate that Fe has higher diffusion rate compared to Ni and

- Co, which again shows similarities to the tracer diffusion coefficients reported in the literature and the intrinsic diffusion coefficients estimated in this study.
- The measured PT diffusion profiles and estimated PT cross interdiffusion coefficients indicate on the diffusional interactions between components and how the interdiffusion fluxes of different components are increased or decreased depending on the diffusion direction of components in different couples.
- Therefore, for the first time, this study indicates that the diffusion coefficients estimated following the newly proposed methods are fundamental in nature and can be correlated with the tracer diffusion coefficients.

**Acknowledgement:**

We would like to acknowledge the financial support from ARDB, India, Grant number: ARDB/GTMAP/01/2031786/M. We acknowledge the support of Hemanth Kumar S and Saswata Bhattacharya of IIT Hyderabad for the help of extracting thermodynamic data from ThermoCalc.